\documentclass[twocolumn,secnumarabic, nobibnotes, aps, prd, 10pt]{revtex4-1}
\usepackage{amsmath,amssymb,bm}
\usepackage{epsfig}
\usepackage{graphicx}
\usepackage{pstricks,pst-node,pst-text,pst-3d}
\usepackage{subfigure}
\usepackage{hyperref}

\newcommand{\avg}[1]{\left\langle #1 \right\rangle}
\newcommand{\vecr}{\bm{r}}

\newcommand{\vk}{\bm{k}}
\newcommand{\x}{\bm{x}}
\newcommand{\y}{\bm{y}}

\newcommand{\ab}{\bar\alpha}
\newcommand{\as}{\alpha_s}

\begin{document}
\title{\bf Inclusive hadron and photon production at LHC in dipole momentum space}

 \author{E.\ A.\ F.\ Basso}
  \email{andre.basso@ufrgs.br}

 \author{M.\ B.\ Gay Ducati}
  \email{beatriz.gay@ufrgs.br}
  \affiliation{Instituto de F\'{\i}sica, Universidade Federal do Rio
                        Grande do Sul, Caixa Postal 15051, 91501-970 --- Porto Alegre, RS,
    Brazil}

 \author{E.\ G.\ de Oliveira}
  \email{egdeoliveira@if.usp.br}
  \affiliation{Instituto de F\'{\i}sica, Universidade de S\~ao Paulo, Caixa Postal 66318, 05314-970 S\~ao Paulo, SP, Brazil}
  \affiliation{Institute for Particle Physics Phenomenology, University of Durham,
Durham, DH1 3LE}

\date{\today}

\begin{abstract}

Using a momentum space model for the dipole scattering amplitude we present an analysis of the saturation effects at LHC energies, describing the data on proton-proton and proton-lead collisions. The model is based on the asymptotic solutions of the Balitsky-Kovchegov equation, being ideal in the saturation domain where the target wave function has a high occupation number. We also make predictions for the nuclear modification ratios on charged hadron and prompt photon production in the forward region, where the high parton density effects are important.

\end{abstract}
\maketitle

\section{Introduction}

The ALICE collaboration at the LHC has recently released \cite{alice_pPb} preliminary data on proton-nucleus collisions, which will be crucial to the understanding of the hadronic matter forming the quark gluon plasma in the case of nucleus-nucleus collisions. This is so because on the $p+Pb$ run the presence of an intricate final state is not expected, being this sort of collision important to study the initial state phenomena and separate these from those intrinsic to the final state. In such a high energy collision, the initial state is expected to be a many body system whose dynamics can be described through the color glass condensate (CGC) formalism (for reviews see \cite{Gelisetal,KovJamal,Weigert}). In this framework, gluon recombination processes should happen in addition to a radiation one, in order to saturate the fast growth of the cross section and thus keep the scattering amplitude unitary. The main aspect of such saturation formalism is the presence of a dynamically generated scale, the saturation scale $Q_s(Y)$, marking when the recombination effects start to become relevant during the evolution on the rapidity variable $Y = \ln(1/x)$ \cite{glr,muelqiu,agl}.  

The simplest way to look for such phenomena is using the Balitsky-Kovchegov (BK) evolution equation for the color dipole scattering amplitude \cite{bk}. It contains all the powers of $\as \log(1/x)$ in the LL${}_\textrm{x}$ approximation, and includes a nonlinear term responsible for the gluon merging in the target wave function. At the moment the numerical solutions of the BK equation are known at NLO accuracy, giving rise to a numerical model of the dipole scattering amplitude \cite{bkrc_num}. Here, however, we will focus on the phenomenology of its asymptotic solutions, which are easily obtained through a mapping of the BK equation onto the Fisher-Kolmogorov-Piscounov-Petrovsky (FKPP) \cite{fkpp} reaction-diffusion equation. 
This statistical formulation allowed us to see the universal behavior of the high energy QCD amplitudes. One knows that the saturation scale increases with a power law of the rapidity in the LO, while it behaves as $\sqrt{Y}$ in the NLO case \cite{mp}. Another striking property of such formulation is related to the geometrical scaling property of the DIS cross sections \cite{gscaling}. Such property, observed in the HERA DIS data, establishes that in the very high energy limit the cross section depends on only one variable, the ratio $Q^2/Q_s^2(Y)$, instead of both $Q^2$ and $Y$ variables separately, making clear the importance of the saturation scale on the observables at high energies. It turns out that such scaling behavior can be seen as the formation of a traveling wave pattern in the solutions of the BK equation, in the analogy with the FKPP equation \cite{mp}. 

Some phenomenology has been done in the last few years using the traveling wave description of the high energy QCD amplitudes, but here we will focus on the AGBS model for the dipole amplitude \cite{agbs}. This model was already used to study the possible presence of the fluctuation effects on the gluon number in DIS data \cite{agbs_fluct}, as well as to investigate the nuclear effects present in the small-$x$ region \cite{betemps_machado09}.   
Recently it has been shown that the AGBS model describes equally well the DIS and inclusive hadron production in $p+A$ and $p+p$ collisions \cite{bgo}. This was made in a new simultaneous fit to HERA data on the proton structure function \cite{h1+zeus} and RHIC data on the $p_t$ distribution of the produced hadrons and pions from the BRAHMS and STAR collaborations \cite{brahms04,star06}. The resulting fit, which is based on the hybrid formalism and uses the parton distribution functions (PDFs) for the hadronic content of the projectile, also describes equally well the LHC data on the hadron yield for $p+p$ collisions, although with large $K$ factors. Such formalism is good to describe the proton fragmentation region in the forward direction, where the nucleus is probed at very small $x$ while the proton's (projectile) values of $x$ are larger; however it is not the best choice in the central rapidity region. Here both hadrons' wave functions are probed at small $x$, so that one can employ the $k_t$-factorization formalism where the colliding hadrons are described in terms of their unintegrated gluon distributions (UGDs), or simply unintegrated dipole distributions in the case of the BK evolution for the scattering amplitudes.

In this work the AGBS dipole model is confronted with the LHC data on single inclusive charged hadron production from $p+p$ reactions using the $k_t$-factorization formalism. It is also applied to proton-nucleus collisions at the LHC, comparing the predictions with the first data on $p+Pb$ measured by the ALICE Collaboration. 
Moreover, the analysis within the hybrid formalism, done in \cite{bgo}, is extended to predict the nuclear modification factors in inclusive hadron and photon production. The latter is known as a good probe to the initial effects, as it does not interact through the strong force with the hadronic content in the final state.

The paper is organized as follows: in Sec. \ref{sec1} we review the LO BK equation in momentum space and its traveling wave solutions giving rise to the AGBS model. Sec. \ref{sec2} is devoted to establishing the formulas and physics behind the $k_t$-factorization formalism used in this work and in Sec. \ref{sec3} we describe the CGC factorization and results for hadron and photon production at the LHC within the hybrid formalism. The final discussion is left for Sec. \ref{sec4}.

\section{BK solutions and the AGBS model}\label{sec1}

The Balitsky-Kovchegov (BK) equation \cite{bk} is a nonlinear evolution equation in the rapidity variable $Y = \ln(1/x)$ for the forward amplitude ${\cal N}(\vecr,Y)$ of a $q\bar{q}$ dipole scattering with coordinates $\vecr=\x-\y$ off a target. It can be derived in Mueller's dipole picture \cite{ahm94} for the high energy scattering, that uses the large $N_c$ limit approximation proposed by 't Hooft in the 1970s and thus guarantees that the evolved dipoles interact independently with the target. When neglecting the impact parameter dependence, one can write the dipole amplitude in momentum space as 

\begin{equation}\label{eq:bk-fourier}
N(Y,k) = \int \frac{d^2\vecr}{2\pi} e^{\imath \vk \cdot \vecr} \frac{{\cal N}(Y,\vecr)}{r^2},
\end{equation}
with $r$ being the transverse size of the interacting dipole, while $k$ denotes its reciprocal transverse momentum. The BK evolution equation for this amplitude reads
\begin{equation}\label{eq:bk_mom}
\partial_Y N(Y,k)=\bar{\alpha}\chi(-\partial_L) N(Y,k)-\bar{\alpha} N(Y,k)^{2},
\end{equation}
where
\begin{equation}\label{eq:kernel}
\chi(\gamma)=2\psi(1)-\psi(\gamma)-\psi(1-\gamma)
\end{equation}
is the BFKL \cite{bfkl} kernel defined in terms of the digamma functions, $\psi(\gamma) = \Gamma^\prime(\gamma)/\Gamma(\gamma)$, while $L=\log(k^2/k_0^2)$, $k_0$ being a fixed soft scale.

The asymptotic solutions of such an equation can be obtained through a map of the BK equation into the Fisher-Kolmogorov-Piscounov-Petrovsky reaction-diffusion equation \cite{fkpp}, known to admit traveling waves as solutions. This implies that the wave fronts have the form $f(x- v_c t)$ for large values of $k$, which translates into the geometrical scaling property of the BK amplitudes \cite{mp}
\begin{equation}\label{eq:Ttail}
\begin{split}
N\left(Y,k\right) \stackrel{k\gg Q_s}{\approx}
 & \left(\frac{k^2}{Q_s^2(Y)}\right)^{-\gamma_c}\log\left(\frac{k^2}{Q_s^2(Y)}\right)\\
 & \times \exp\left[-\frac{\log^2\left(k^2/Q_s^2(Y)\right)}{2\bar{\alpha}\chi''(\gamma_c)Y}\right],
\end{split}
\end{equation}
where the saturation scale is defined as
\begin{equation}\label{eq:satscal}
Q_s^2(Y) = Q_0^2\exp\left( \lambda Y - \frac{3}{2\gamma_c}\log Y \right),
\end{equation}
with $\lambda  = \ab \chi(\gamma_c)/\gamma_c$ 
measuring how fast the amplitudes reach the saturated domain.

Within this traveling wave description of the high energy QCD amplitudes, a model for the dipole scattering amplitude was proposed, which describes the infrared behavior as a Fourier transform of a step function $\Theta(r Q_s - 1)$, demanding unitary amplitudes in the coordinate space. Thus, in this region the amplitude is written as
\begin{equation}\label{eq:Tsat}
N\left(Y,k\right) \stackrel{k\ll Q_s}{\approx} c - \log\left(\frac{k}{Q_s(Y)}\right),
\end{equation}
where $c$ is a constant. The AGBS dipole model \cite{agbs} interpolates analytically between these two regions through
\begin{equation}\label{eq:agbs}
N(Y,k) = \left[ \log\left(\frac{k}{Q_s} + \frac{Q_s}{k}\right) + 1\right](1-e^{-T_\textrm{dil}}),
\end{equation}
with
\begin{equation}\label{eq:agbs_dil}
T_\textrm{dil} = \exp\left[ -\gamma_c\log\left(\frac{k^2}{Q_s^2(Y)}\right) - \frac{L_\textrm{red}^2 - \log^2(2)}{2\ab\chi^{''}(\gamma_c)Y}\right],
\end{equation}
and
\begin{equation}\label{eq:agbs_Qsat}
L_\textrm{red} = \log\left( 1 + \frac{k^2}{Q_s^2(Y)} \right), \qquad Q_s^2(Y) = k_0^2\,e^{\lambda Y}.
\end{equation}

Once it describes the evolution for a quark-antiquark amplitude, the BK solutions are written in the fundamental representation for the Wilson lines $U$ of the gauge fields in the target
\begin{equation}
N_F(Y,r) = 1 - \frac{1}{N_c} Tr\avg{U^\dagger(0)U(\vecr)}_Y.
\end{equation}
The adjoint amplitude for gluons required in the $k_t$-factorization formalism can be written, in the large $N_c$ limit, as
\begin{equation}\label{eq:adj_lnc}
N_A(Y,r) = 2N_F(r,Y) - N_F^2(r,Y).
\end{equation}

In the case of the AGBS dipole model, however, it was shown that one can get the adjoint amplitude $N_A$ by a rescaling of the saturation scale entering the fundamental one \cite{bgo}. In other words, to get $N_F$ one can make the replacement $Q_s^2 \rightarrow (C_A/C_F)Q_s^2$, with $C_A/C_F = 9/4$, in the expression for $N_A$, in order to make both the hadron production and DIS amplitude compatible. As shown in Ref. \cite{bgo}, this is a good approximation of the usual expression (\ref{eq:adj_lnc}) within the framework of  the AGBS model, giving a gluon saturation momentum higher than in the gluon case. This is what one would expect in the large $N_c$ limit where a gluon is seen as two quarks. The resulting forms for both adjoint and fundamental amplitudes in momentum space are depicted in Fig. \ref{fig:agbs_amp}.

\section{Hadron production from $k_t$-factorization}\label{sec2}

In the regime of very high energies, for $\sqrt{s}\gg Q_s \geq p_t$, the dominant process contributing in the high energy cross section is the gluon production via gluon-gluon fusion and the subsequent fragmentation of the produced gluon. When the scattering process occurs between a dilute projectile and a dense target, with a large occupation number, the cross section for production of a gluon jet with transverse momentum $q_t$ can be described by the $k_t$-factorization formalism \cite{kt-fact}
\begin{equation}\label{eq:kt_fact}
\begin{split}
\frac{d\sigma^{A+B\rightarrow g} }{dy d^2q_t} & =  K \frac{2}{C_F q_t^2}\int^{q_t} \frac{d^2k_t}{4}  \as(Q) \\
& \times  \varphi\left(  x_1, \frac{|q_t + k_t|}{2} \right)\,\varphi\left(  x_2, \frac{|q_t - k_t|}{2}\right),
\end{split}
\end{equation}
where $x_{1,2} = (q_t/\sqrt{s})e^{\pm y}$ are momentum fractions of the incoming gluons and $C_F = (N_c^2 - 1)/2N_c$ is the Casimir for the fundamental representation.

Even though the model we used is based on the LO BK solutions, it incorporates phenomenologically some properties of the NLO dipole dynamics, such as the saturation exponent, that is expected to be reduced from $\lambda \sim 0.9$ in LO models to $\lambda \approx 0.2-0.3$ from the data. Thus, concerning the argument of the strong coupling constant in (\ref{eq:kt_fact}), instead of using a fixed value we allow for the running of $\as(Q)$ within the LO prescription for three light quark flavors

\begin{equation}
\as(Q^2) = \frac{12\pi}{27\log\frac{Q^2}{\Lambda_{\text{QCD}}^2}}\,,
\end{equation}
with the maximum value of momentum of the UGDs as the scale; {\it i.e.}, we use $Q = \max\{|q_t+k_t|/2,|q_t-k_t|/2\}$, and set $\Lambda_{\text{QCD}}^2 = 0.05$ GeV${}^2$.

\begin{figure}[t!]
\centering
\subfigure[]{
\scalebox{1}{\includegraphics{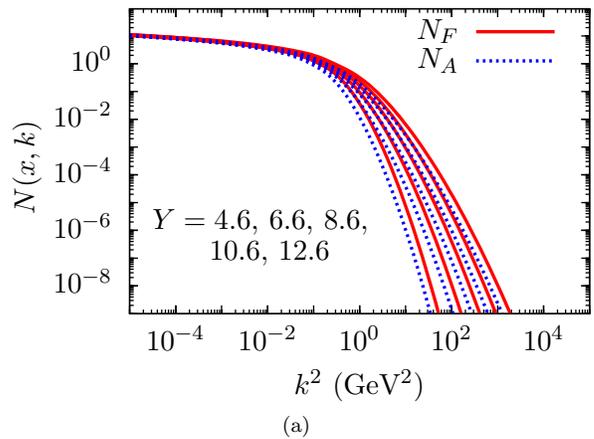}}
\label{fig:agbs_amp}
}
\centering
\subfigure[]{
\scalebox{1}{\includegraphics{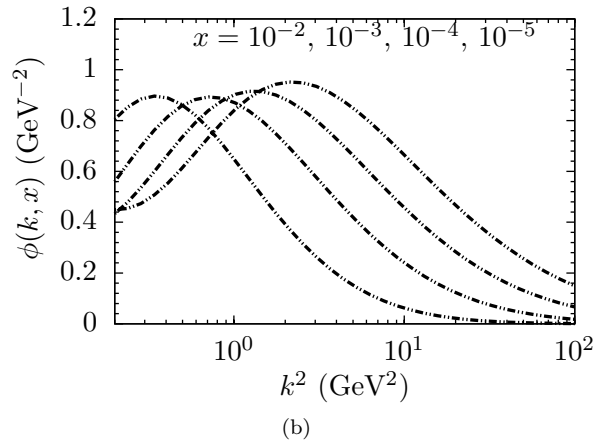}}
\label{fig:agbs_ugd}
}
\caption{(a) AGBS amplitude in momentum space and (b) the respective unintegrated gluon distribution of a proton.}
\label{fig:amp_ugd}
\end{figure}

In the large $N_c$ limit, the unintegrated dipole gluon distribution in either of the two colliding hadrons can be related to the dipole scattering amplitude through \cite{Braun}

\begin{equation}\label{eq:ugd}
\varphi\left(x, k_t \right) = \frac{N_c\,S_t}{2\pi^2\as(k_t)} k^2 F_g(x,k_t),
\end{equation}
where $N_c$ is the number of colors and $S_t$ is the transverse area of the interacting hadrons (nuclei).
The quantity $F_g(x,k_t)$ stems from the adjoint dipole scattering amplitude. As one can see in Eq. (\ref{eq:bk-fourier}), the BK amplitude in momentum space is the subject of a slightly different Fourier transform, but this can be overcome using \cite{Braun}
\begin{equation}
F_g(x,\vk) = \frac{1}{2\pi}(\nabla_{\vk}^2 N_A(x,\vk) + \delta^{(2)}(\vk)),
\end{equation}
where the delta function can be neglected as the unintegrated dipole gluon distribution (\ref{eq:ugd}) vanishes for $k_t=0$. Such prescription was successfully used with the AGBS model to describe simultaneously the HERA data on the proton structure functions and the inclusive hadron yield of charged hadrons at the RHIC \cite{bgo}. The functional form of the UGD is shown in Fig. \ref{fig:agbs_ugd}.

\subsection{Results}\label{sec2a}

The transverse momentum distribution of the produced charged hadrons is given by
\begin{equation}\label{eq:had-yield}
\frac{dN_{ch}}{d\eta d^2p_t} = \frac{J[\eta]}{\sigma_{nsd}} \int \frac{dz}{z^2} \frac{d \sigma^{A+B\rightarrow g}}{dy d^2q_t} D_h(z=p_t/q_t, \mu)\,,
\end{equation}
where $J[\eta]$ is the Jacobian for the transformation of the rapidity $y$ and the measured pseudorapidity $\eta$ variables 
\begin{equation}
y(\eta,p_t) = \frac{1}{2}\ln\left(  \frac{\sqrt{m^2 + p_t^2\cosh^2\eta} + p_t\sinh\eta}{\sqrt{m^2 + p_t^2\cosh^2\eta} - p_t\sinh\eta} \right)\,,
\end{equation}
with $m$ denoting the hadron mass. 
In (\ref{eq:had-yield}), $D_h(z=p_t/k_t,\mu)$ stands for the fragmentation function of the produced gluon into hadrons, for which the LO KKP model \cite{kkp} is used, at the scale $\mu=p_t$ of the hadron and with $z_\textrm{min}=0.05$, as demanded from the momentum sum rule. 
To avoid the divergences appearing for small values of $p_t$ in the $k_t$-factorization formula (\ref{eq:kt_fact}), we make the change $p_t \rightarrow \sqrt{m^2 + p_t}$, where the hadron mass $m$ has the same value as that used in the Jacobian relating the rapidity and pseudorapidity variables.

\begin{figure}[t!]
\large
\begin{center}
\scalebox{0.9}{\includegraphics{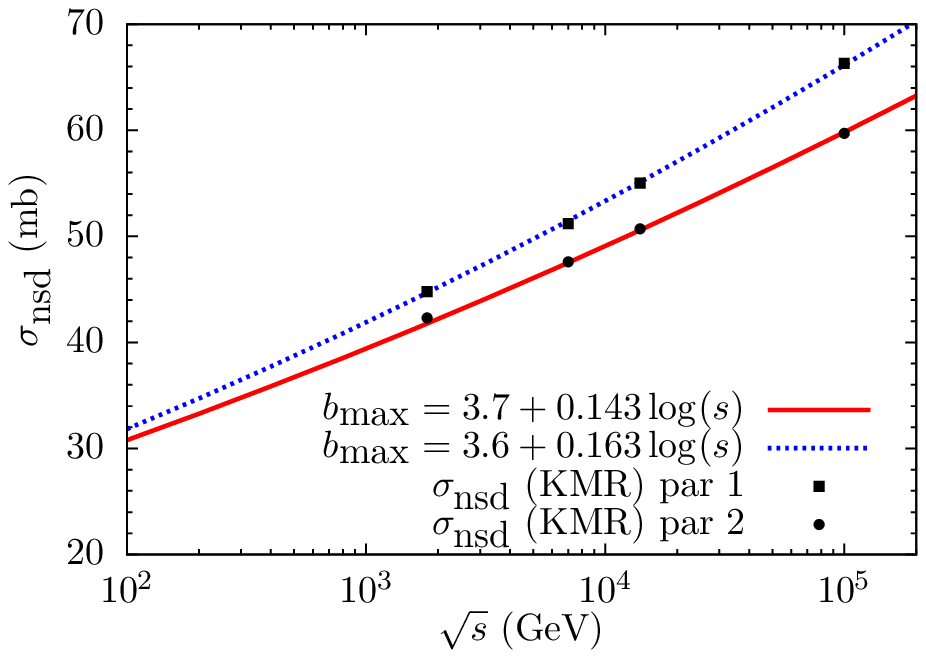}}
\caption{Comparison between the results from the different parameter sets of the KMR model \cite{kmr_2011} for the nonsingle diffractive cross section $\sigma_{nsd}$ and the prescription $\sigma_{nsd} = \pi b_\textrm{max}^2$, where $b_\textrm{max} = b_0 + C \log(s)$. }
\label{fig:sig_nsd}
\end{center}
\end{figure}

\begin{figure}[t!]
\large
\begin{center}
\scalebox{1.0}{\includegraphics{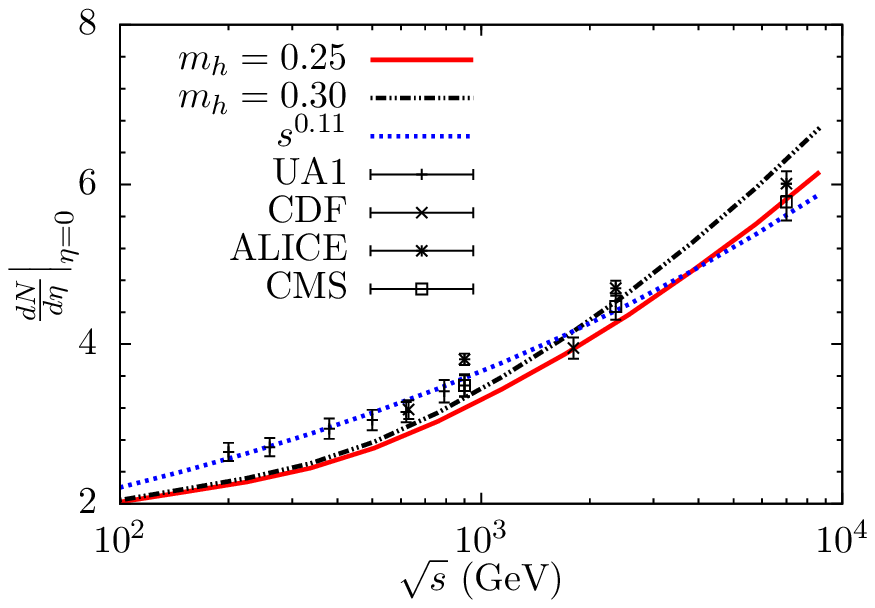}}
\caption{Central rapidity distribution of hadrons in $p(\bar{p} + p)$ collisions at different energies. The prescription $b_\textrm{max} = b_0 + C \log(s)$ fairly describes the data for the values $b_0 \approx 0.15$ and $C \approx 0.2$. For comparison, the power law curve that fits the data, which were taken from \cite{ua1,cdf,alice_eta0,cms_eta0}, is also plotted.}
\label{fig:cent_rap}
\end{center}
\end{figure}

The nonsingle diffractive cross section $\sigma_{nsd}$ is, in principle, model dependent and should be taken from soft models such as in Refs.\ \cite{kmr_2011, Gotsman_2008}. However, the physical meaning of $\sigma_\textrm{nsd}$ being the interaction area, we can model it, as done for instance in Refs.\ \cite{Levin-rezaeian-2010,tv}. Following \cite{tv}, an energy-dependent interaction radius $b_\textrm{max} = a + b \log(s)$ is introduced, such that $\pi b_\textrm{max}^2$ mimics the nonsingle diffractive cross section. Two alternatives to fix the parameters $a$ and $b$ are used here: to reproduce the results of the model \cite{kmr_2011}, that includes both soft and hard contributions, which is shown in Fig. \ref{fig:sig_nsd} for two different parameter sets obtained in such model; and to describe the central production of charged hadrons $dN/d\eta |_{\eta = 0}$ at different values of the center of mass energy. The best agreement of the AGBS model with the data implies $a=0.1$ and $b=0.198$ and the results of this choice are shown in Fig. \ref{fig:cent_rap}. We will show our results for the charged hadron yield using the first choice for $\sigma_\text{nsd}$, noting that following the KMR model \cite{kmr_2011} the central rapidity data cannot be described for a broad range of $\sqrt{s}$ in the framework of the AGBS model. Thus, the last choice seems to be the best one to fix the normalization of our calculation. Being an overall normalization, this choice for the nonsingle diffractive cross section allows for a $K$ factor of the order of unity, even though one does not expect such values for a LO calculation.

The AGBS model was fitted to HERA data considering the small-$x$ region data, $x<0.01$, so that quark corrections to the CGC formalism could be ignored. In proton-nucleus or proton-proton collisions, however, one has to consider corrections due to large-$x$ effects, which were introduced in our calculation through the relation \cite{GSV}
\begin{equation}\label{eq:qcorr}
\varphi\left(k_t, x \right) = \left( \frac{1-x}{1-x_0} \right)^\beta \left( \frac{x_0}{x} \right)^{\lambda_0} \varphi\left(k_t, x_0 \right)\,,
\end{equation}
where $\beta=4$ and $\lambda_0$ can vary from 0 to 0.2 in the proton-proton case.

\normalsize

\normalsize

\begin{figure}[t!]
\centering
\subfigure[]{
\scalebox{1.05}{\includegraphics{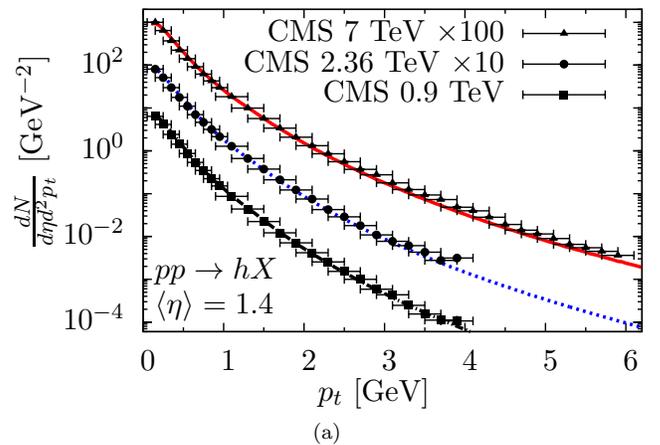}}
\label{fig:Ptdist_lhc}
}
\centering
\subfigure[]{
\scalebox{1.2}{\includegraphics{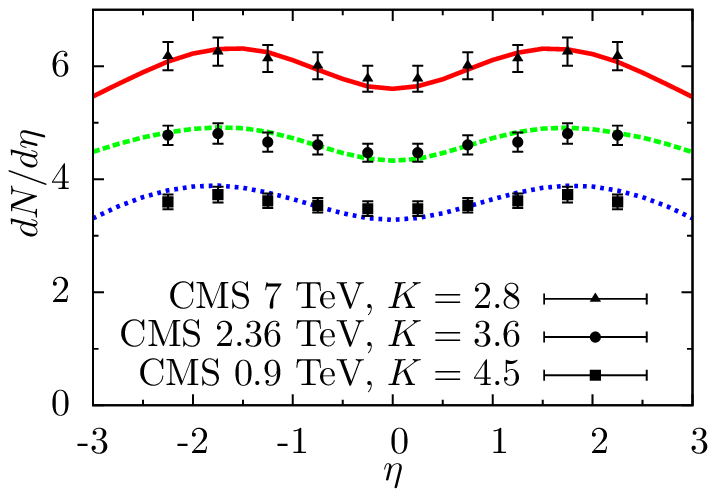}}
\label{fig:etadist_lhc}
}
\label{fig:LHC-yield}
\caption{$p_t$ (a) and $\eta$ (b) distributions of the LHC charged hadron yield for $p+p$ collisions at 0.9, 2.36 and 7 TeV \cite{cms-1,cms-2}. AGBS parameters of the simultaneous fit \cite{bgo} of AGBS to RHIC \cite{brahms04,star06} and HERA  \cite{h1+zeus} data. The normalization used follows $\pi b_\textrm{max}^2$, with $b_\textrm{max}$ taken from Fig. \ref{fig:sig_nsd} and the $K$ factors are the same for both observables. We note that using the normalization from Fig. \ref{fig:cent_rap} we could describe the data with a $K$ factor close to 1.}
\end{figure}

In Figs. \ref{fig:Ptdist_lhc} and \ref{fig:etadist_lhc} are shown the transverse momentum and pseudorapidity distributions of charged hadrons at the LHC for 0.9, 2.36 and 7 TeV $p+p$ collisions, which present a good description of the data. Concerning the $K$ factors we got a better description of the LHC data in comparison with the previous work using the hybrid formalism \cite{bgo}. This result is expected once the $k_t$-formalism is better suited to deal with central rapidity data as that measured by the CMS Collaboration, while the hybrid formalism describes the forward hadron fragmentation region.
It is worthwhile to note that such formalism needs corrections due to inelastic scatterings between the colliding particles \cite {ak} that should play an important role at the midrapidity region and at high hadron $p_t$, while in the forward regions, the inelastic piece has a very small contribution to the production cross section \cite{amir_jamal_2011}. Such corrections, however, are also included in the full one loop calculation to the inclusive hadron production recently done in \cite{nlo_ihp,beuf_f2nlo}, and we leave for a future study the phenomenological applications of these observables within the traveling wave method of QCD.

\begin{figure}[t!]
\begin{center}
\scalebox{1.1}{\includegraphics{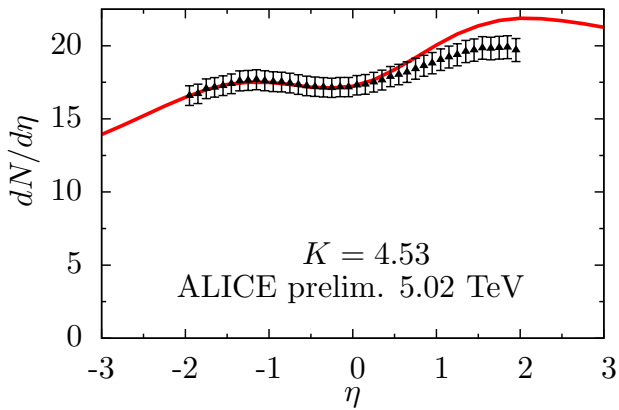}}
\caption{Pseudorapidity distribution of the charged hadrons produced in $p+Pb$ collisions at LHC \cite{alice_pPb}. 
}
\label{fig:alicepPb}
\end{center}
\end{figure}

In Fig. \ref{fig:alicepPb} we show the pseudorapidity distribution against the recent preliminary data on proton-lead collisions measured by the ALICE Collaboration \cite{alice_pPb}.
The nuclear saturation scale is modeled through $Q^2_{s,A} = A_\textrm{eff}^{1/3} Q^2_{s,p}$, where $A_\textrm{eff}=20$ for lead targets. We have also checked the prescription of Ref. \cite{asw}, for which $Q^2_s,A = (A\,R_A^2/R_p^2)$ (with $R_A = 1.12\,A^{1/3} - 0.86\,A^{-1/3}$ fm), and the description is similar. The last one was already used with the AGBS model for the nuclear ratio $R(A/B) = BF_2^A/AF_2^B$ \cite{betemps_machado09}, with a good description of the shadowing region.

Looking at Fig. \ref{fig:alicepPb} one can see that the proton forward region, for negative pseudorapidity, is well described by the model. This is expected, once the small-$x$ ($x_2\ll 1$) effects of the nucleus are encoded in this region. Regarding the positive rapidity range of the nucleus, it should be important to include a better prescription of the nuclear geometry. An implementation of the impact parameter in the model might improve the description of the nuclear data, once both the large-$x$ nuclear effects, like EMC and Fermi motion, and the small-$x$ nuclear shadowing effects are embedded into the saturation scale.

The result presented here for the proton-nucleus collisions is similar to that of the IP-Sat saturation model \cite{ip_sat_lhc}, mainly in the case of LHC collisions, even though we do not use any term to include the impact parameter dependence on the saturation scale. Comparing our predictions with the data released recently by the ALICE Collaboration \cite{alice_pPb} one can see that the ratio 

\begin{equation}
\frac{dN/d\eta|_{\eta=-2}}{dN/d\eta|_{\eta=2}}
\end{equation}
is almost the same as the IP-Sat model one, for which the ratio is $1.32$ against $1.33$ for the AGBS model (the KLN \cite{kln_lhc} and the rcBK \cite{rc_bk_lhc} models show the ratios $1.38$ and $1.42$, respectively). Thus one can see that concerning the rapidity evolution, even a LO model such as the AGBS one can describe quite well the present data, despite the large-$x$ regions we already discussed. The same is not true regarding the transverse momentum distribution, where the LO dipole models would fatally fail to describe the large $p_t$ region; once in this region the inclusion of higher order terms on the dipole evolution, or even including a virtuality evolution, is mandatory. One should still stress that all these saturation models are not close to the value $1.19$ of the experiment \cite{alice_pPb}, maybe signalizing that for the LHC data on single inclusive particle spectra a $Q^2$ evolution is important as well.

We can also use the central rapidity data of produced hadrons to constrain our parametrization for the nonsingle diffractive cross section, as shown in Fig. \ref{fig:cent_rap}, and thus decrease the values of the $K$ factors. Indeed, we checked that it is possible to describe the data with a $K$ factor of order 1 for the entire LHC energy spectrum. The RHIC data imply a little larger value as a consequence of the normalization shown in Fig. \ref{fig:cent_rap}, that in the range of RHIC energy underestimates the data.

\normalsize

\section{The CGC factorization for hadron and photon production}\label{sec3}

In this section we explore the particle production within the CGC factorization for high energy scattering. 
Unlike the $k_t$ factorization formalism, the CGC factorization deals with particle production in the fragmentation region of the colliding hadrons. The CGC formalism allows for a factorization in the high energy limit that considers the multiple scatterings of a projectile parton with a dense target. The formalism is explained in \cite{cgc-hadron, cgc-photon} and considers the processes $q(g) T \rightarrow \dots X$, where the dots indicate a particular observable. In what follows this formalism will be used to study hadron and photon production.

\subsection{Hadron production at the LHC}\label{sec3a}

\begin{figure}[t!]
\begin{center}
\scalebox{0.9}{\includegraphics{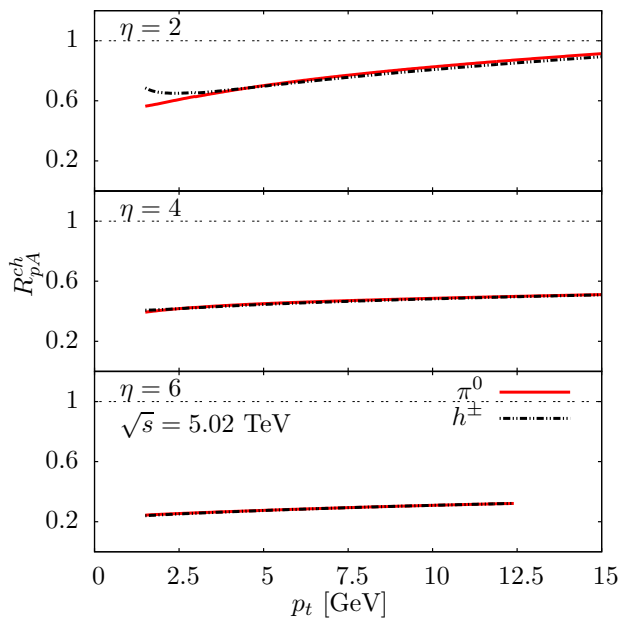}}
\caption{Predictions of the AGBS model to the nuclear modification ratio $R_{pA}$ to the hadron production at $\sqrt{s}=5.02$ TeV. 
}
\label{fig:Rpa-had-lhc}
\end{center}
\end{figure}
\normalsize

The hadron production within the CGC formalism is described through the hybrid factorization of \cite{dhj,ak}, which reads
\begin{equation}\label{eq:had_yield}
\begin{split}
\frac{dN}{dy_h\,d^2p_t} = & \frac{K}{(2\pi)^2}\int_{x_F}^1 \frac{dz}{z}\\
                          \times&  \left[x_1\,f_{q/p}(x_1,p_t^2)\tilde{N}_F\left(\frac{p_t}{z},x_2\right)D_{h/q}\left( z, p_t^2\right)\right.\\
& \left. + x_1\,f_{g/p}(x_1,p_t^2)\tilde{N}_A\left(\frac{p_t}{z},x_2\right)D_{h/g}\left( z, p_t^2\right) \right] \\
+ & \,\delta^\text{inel}\,,
\end{split}
\end{equation}
where $\tilde{N}_{A,F}$ denote the adjoint and fundamental amplitudes for the gluon and quark multiple interactions with the target and $\delta^\text{inel}$ contains the inelastic corrections of \cite{ak}. The function $f_{q,g/p}(x_1,p_t^2)$ stands for the projectile PDFs, which were taken from the CTEQ6 parametrization \cite{cteq6} at LO with $p_t^2$ as scale. The momentum fractions are defined as follows:

\[x_F=\sqrt{m_h^2 + p_t^2}\exp(\eta_h)/\sqrt{S_{NN}}\approx p_t\exp(y_h)/\sqrt{S_{NN}}\]
and
\[x_2 = x_1\exp(-2y_h), \qquad x_1 = x_F/z,\]
where one uses the approximation $\eta_h \approx y_h$, valid for light hadrons. 

To avoid the uncertainties due to higher order corrections in our calculation, only ratios of observables are used, canceling the $K$ factors and minimizing errors. Here the inelastic corrections on (\ref{eq:had_yield}) are also neglected.
The predictions of the AGBS model to the nuclear modification ratio are obtained from 
\begin{equation}
R_{pA}^{h} = \frac{dN^{p A \rightarrow h X}}{d^2p_t d\eta} \Biggl/ \frac{dN^{p p \rightarrow h X}}{d^2p_t d\eta}  \Biggl/ N_\text{coll}\,.
\end{equation}
The number of binary collisions, $N_\text{coll}$, comes from the Glauber Monte Carlo approach, and $N_\text{coll} = 6.9$ was used for minimum-bias proton-lead collisions at $\sqrt{s} = 5.02$ TeV \cite{denterria_03}. The results, depicted in Fig. \ref{fig:Rpa-had-lhc}, show a strong suppression at forward rapidities, that is almost constant even at high hadron transverse momentum. However, when the hadrons are produced more centrally, the ratio increases at large $p_t$. This might be related to the fact that the calculation here employed misses the NLO corrections to the dipole evolution, which is supported by the recent results with the rcBK model \cite{amir_2013}, where such increase at high $p_t$ does not happen, even for central ($\eta=0$) hadron production.

\subsection{Prompt photon production}\label{sec3b}

Within the CGC formalism, the inclusive photon production can be obtained from the semi-inclusive process $qT\rightarrow \gamma q X$ after integration over the final quark momentum, as done in \cite{aj-photon}.  The cross section for prompt photon production in a proton(deuteron) collision is given by
\begin{equation}\label{eq:promt-photon}
\frac{d\sigma^{pT \rightarrow \gamma(k) X}}{d^2\bm{k}_t d\eta_\gamma} = \int_{x_q^\text{min}}^1 d x_{q,\bar{q}} f_{q,\bar{q}}(x_{q,\bar{q}}, k_t^2)  \frac{d\sigma^{q(p) T \rightarrow \gamma(k) X}}{d^2\bm{k}_t d\eta_\gamma}\,,
\end{equation}
where $T$ denotes the target, $f_{q,\bar{q}}(x_{q,\bar{q}},k_t^2)$ is the PDF for a quark (antiquark) with momentum fraction $x_q$ of the projectile and a summation over quark flavors is understood. Here again the LO CTEQ6 distribution \cite{cteq6} for the quark (antiquark) content in the projectile was used. The partonic cross section for photon production by a quark (antiquark) can be defined as the sum of both the fragmentation photons and the direct photon contributions. This was done in \cite{aj-photon} and the final invariant cross section reads

\begin{equation}\label{eq:inc-photon}
\frac{ d\sigma^{q(p) T \rightarrow \gamma(k) X} }{d^2\bm{k}_t d\eta_\gamma} =  \frac{ d\sigma^\text{Fragmentation} }{d^2\bm{k}_t d\eta_\gamma} + \frac{ d\sigma^\text{Direct} }{d^2\bm{k}_t d\eta_\gamma},
\end{equation}
where
\begin{equation}\label{eq:phot-frag}
\frac{ d\sigma^\text{Fragmentation} }{d^2\bm{k}_t d\eta_\gamma} = \frac{1}{(2\pi)^2} \frac{1}{z} D_{\gamma/q}(z,Q^2) N_F(x_g, k_t/z),
\end{equation}
and
\begin{equation}\label{eq:phot-dir}
\frac{ d\sigma^\text{Direct} }{d^2\bm{k}_t d\eta_\gamma} = \frac{e_q^2 \alpha_\text{em}}{\pi (2\pi)^3} z^2 \left[ 1 + (1-z)^2 \right]\frac{1}{k_t^4} \int^{k_t^2} d^2 \bm{l}_t l_t^2 N_F({\bar{x}}_g, l_t).
\end{equation}

The leading order quark-photon fragmentation function $D_{\gamma/q}(z,Q^2)$ has the form \cite{owens}
\begin{equation}\label{eq:qphot-ff}
D_{\gamma/q}(z,Q^2) = \frac{e_q^2 \alpha_\text{em}}{2\pi} \frac{1 + (1-z)^2}{z} \log\left( \frac{Q^2}{\Lambda^2} \right)\,.
\end{equation}

The momentum fractions entering the production cross section (\ref{eq:inc-photon}) at a center of mass energy $\sqrt{s}$ were defined in \cite{aj-photon} as follows:

\begin{equation}
x_g = \frac{k_t^2}{z^2 x_q s} = x_q e^{-2\eta_\gamma},
\end{equation}
\begin{equation}
\bar{x}_g = \frac{1}{x_q z}\left[ \frac{k_t^2}{z} + \frac{(l_t - k_t)^2}{1-z} \right],
\end{equation}
\begin{equation}
z = \frac{k_t^2}{x_q \sqrt{s}}\,e^{\eta_\gamma} = \frac{x_q^\text{min} }{x_q}, \quad \text{with} \quad x_q^\text{min} = z_\text{min}= \frac{k_t}{\sqrt{s}}\,e^{\eta_\gamma}.
\end{equation}

\begin{figure}[t!]
\begin{center}
\scalebox{0.95}{\includegraphics{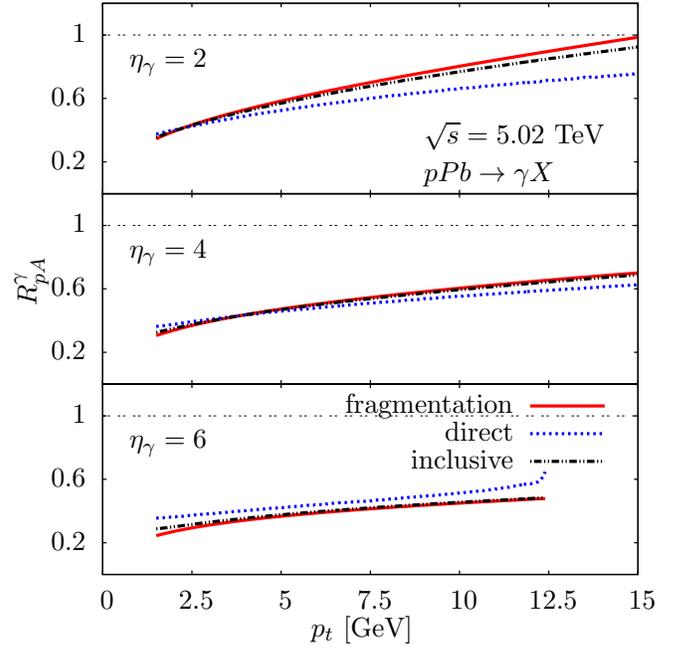}}
\caption{Predictions of the AGBS model to the nuclear modification ration $R_{pA}^\gamma$ for inclusive photon production at $\sqrt{s}=5.02$ TeV. 
}
\label{fig:Rpa-lhc}
\end{center}
\end{figure}

Once again there is the use of ratios to reduce uncertainties, setting $K=1$ throughout this section. The nuclear modification ratio is defined analogously to the hadron production case, and the results for such observable are shown in Fig. \ref{fig:Rpa-lhc}. It is noted that within the AGBS frame, the direct photons dominate over the fragmentation one in the forward region. It can be stressed, however, that no isolation cut on the photon transverse momentum was used, meaning that the fragmentation photons have contamination from the decaying of hadrons into photons. In a more careful future comparison with available data, this isolation should be included.

\normalsize

The AGBS predictions for the ratio of inclusively produced photons to the neutral pions were also investigated. The ratio is defined as

\begin{equation}
\frac{\gamma^\text{inclusive}}{\pi^0} =  \frac{d\sigma^{p T \rightarrow \gamma X}}{d^2 p_t^\gamma d\eta^\gamma} \Biggl/ \frac{d\sigma^{p T \rightarrow \pi^0 X}}{d^2p_t^\text{h} d\eta^\text{h}},
\end{equation}
where the pion production is described through (\ref{eq:had_yield}) and the photon by (\ref{eq:inc-photon}). 
The ratio was calculated for proton-proton and proton-lead collisions at $\sqrt{s}=5.02$ TeV, as a function of transverse momentum and rapidities of the produced particles (pions or photons).

\begin{figure}[t!]
\begin{center}
\scalebox{0.75}{\includegraphics{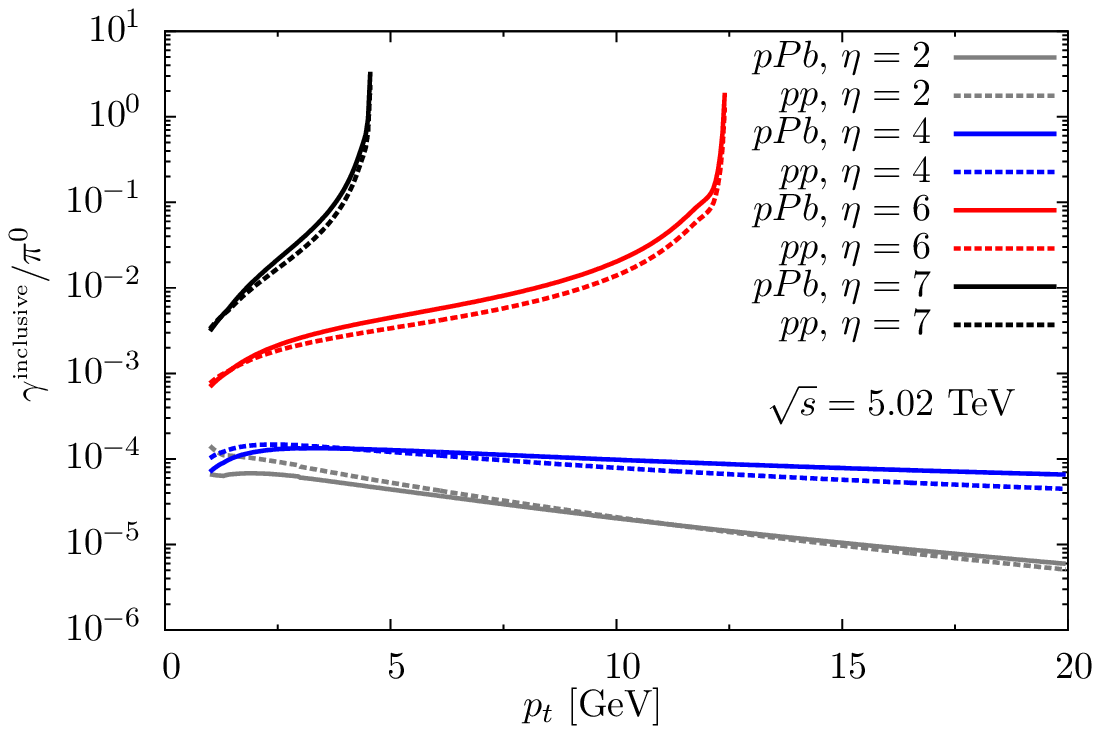}}
\caption{Ratio for the inclusive production of photons and neutral pions at $\sqrt{s}=5.02$ TeV, for different values of pseudorapidity. 
}
\label{fig:gamma-pi}
\end{center}
\end{figure}

The results are illustrated in Fig. \ref{fig:gamma-pi}, showing the ratio is smaller than 1 for a large rapidity region, meaning that the neutral pion production dominates over the inclusive photon production. This occurs since in the CGC approach defining the photon production cross section (\ref{eq:inc-photon}), the photon is radiated by a quark before or after the interaction with the dense target field. On the other hand, the pion production can be generated also from gluons, and these dominate at the midrapidity region at LHC energies. However, in the very forward region, $\eta \sim 7$, the momentum fraction is close to 1 and the quark contribution to the cross section is the main one. Thus, one can see the importance in measuring these observables at forward rapidities, allowing a deep investigation of the CGC factorization here employed, as well as the distinct saturation models. Moreover, it is known that the collinear approach of PDF for the photon production results in a weaker suppression of the nuclear modification factor \cite{aj-photon}. Thus, the forward rapidity region data might also help to disentangle the collinear approach from that $k_t$ factorized embedded in the dipole saturation models.  
This ratio could be used, in the case of the AGBS model, to evaluate the modeling of the nuclear saturation scale, exploring the slight differences for proton and lead targets in a relatively large region of transverse momentum.

\normalsize

\section{Discussion}\label{sec4}

The inclusive hadron production was calculated for the LHC energies within the $k_t$-factorization formalism using a saturation based unintegrated gluon distribution. The same observable was previously studied in Ref. \cite{bgo}, within the hybrid formalism \cite{dhj} for the production cross section, in a global fit that constrained the parameters of the model to both the HERA data on the proton structure function and the inclusive hadron yield at the RHIC. Comparing both approaches it is observed that, even though these formalisms are designed to describe distinct kinematical regions of the data $--$the hybrid one for the forward region and the $k_t$ factorized at the central rapidity region$--$ the results are very similar except for the values of the $K$ factors normalizing all the uncertainties of the model in the multiplicity cross sections.
Indeed, the $K$ factors gotten in this work are smaller compared with the hybrid formalism, as expected once we are dealing with the central data of the CMS experiment \cite{cms-1,cms-2}. Thus a detailed analysis of the region of validity for each formalism is crucial for the study of saturation physics at the LHC, once such phenomena can be accessed in distinct kinematical regions of the detector, depending on the observable and kinematical cuts used. 

The large nucleus effects were studied using the recently measured proton-lead collisions at the LHC. In this case the AGBS model can describe quite well the proton fragmentation region where the nuclear small-$x$ effects are present. The nucleus fragmentation region lacks the implementation of large-$x$ effects on the nuclear target wave function. All in all, one can see that a LO dipole evolution can still describe the present data on the pseudorapidity distribution. 

We have also made predictions for the nuclear modification factors for charged hadron and photon production. The AGBS model shows the expected suppression on the forward rapidities that is observed in saturation models; in the opposite way the collinear nuclear PDFs do not show such behavior. The ratio of inclusive production of photons against pions was also calculated, and the result is slightly different from other dipole models. One sees that the measurement of these observables in the forward rapidity region could help to discriminate between the saturation physics and the collinear one, as well as to constrain the dipole models in the literature.

Concerning the AGBS model and its limitations we note that it describes quite well the data on $p+p$ collisions, but does not incorporate the large-$x$ physics present in the nucleus fragmentation region, when applied to $p+A$ cases. A way to introduce such effects in the model might be the study of its whole transverse dependence. Being a transverse-momentum-dependent gluon distribution function, it should be important to employ this model on semi-inclusive observables probing deeply the transverse momentum on the initial state. Another source of information would come from multiparton correlations, once on those processes a more complex color structure may be probed.
Moreover, the BK equation is obtained using perturbative techniques and the information on its impact parameter dependence is rather a phenomenological issue, as it is a long range effect and should not be treated perturbatively. Thus it might be interesting to model such dependence with parameters constrained by diffractive DIS data, as done for instance in \cite{bs} for inclusive DIS and exclusive vector meson production, so that the dipole model could be used to give a better description of the transverse plane on nuclear reactions.

\section*{Acknowledgements}
The authors thank Y. Kovchegov and C. Marquet for very fruitful discussions.
This work is supported by CNPq (Brazil). E.G.O. is supported by FAPESP (Brazil) under Contract 2012/05469-4.

\bibliographystyle{unsrt}
 
\end{document}